\documentstyle[preprint,prl,aps]{revtex}
\begin{document}
\draft

\title{Noise and dynamical pattern selection}

\author{Douglas A. Kurtze\cite{email}}

\address{Department of Physics, North Dakota State University, SU Station,
Box 5566, \\ Fargo, North Dakota 58105-5566}

\date{November 14, 1995}

\maketitle

\begin{abstract}
In pattern forming systems such as Rayleigh-B\'enard convection or
directional solidification, a large number of linearly stable, patterned
steady states exist when the basic, simple steady state is unstable.  Which
of these steady states will be realized in a given experiment appears to
depend on unobservable details of the system's initial conditions.  We show,
however, that weak, Gaussian white noise drives such a system toward a
{\it preferred wave number\/} which depends only on the system parameters and
is independent of initial conditions.  We give a prescription for calculating
this wave number, analytically near the onset of instability and numerically
otherwise.
\end{abstract}


The classic problem of pattern selection is that of predicting which of a
large number of available steady states a system will ultimately reach under
given experimental conditions.  In a typical example, that of directional
solidification, there is a simple steady state of the system in which the
solidification front is planar and advances into the melt at a constant
speed.  By varying a control parameter, one reaches a regime in which this
state is linearly unstable against all perturbations with wave numbers in a
given interval.  In this regime the front settles into a spatially periodic,
cellular shape.  Ample evidence exists that there is one such patterned
steady state with each wave number $q$ in the interval of instability.
Moreover, there is a finite subrange of wave numbers for which the cellular
steady states are themselves linearly stable.  The pattern selection question
is then this:  Into which of these patterned states will the system
restabilize in a given experiment?

The answer to this question appears to be that the final wave number depends
not only on the system parameters, but also on the details of the initial
conditions from which the system evolves.  Since these details cannot be
observed in practice, the final wave number is not reproducible.  We will
argue in this Letter, however, that among the possible steady state wave
numbers there is one which is {\it preferred}, in the sense that subjecting
the system to weak, Gaussian white noise drives it toward that wave number,
and in the long-time limit that wave number is overwhelmingly more probable
than any other.  We will show how this preferred wave number can be
calculated for one-dimensional systems.

Noise effects on pattern-forming systems have been studied by many authors,
although few have considered the role of noise in readjusting the wave number
of an established, periodic pattern.  Deterministic evolution from random
{\it initial conditions\/} has been studied in amplitude equations
\cite{1,2,3}, and noise effects on the initial stages of pattern formation
has been investigated in B\'enard convection \cite{4}, dendritic growth
\cite{5,6}, and cellular \cite{7} and dendritic \cite{8} arrays in
directional solidification.  Several numerical studies of the Swift-Hohenberg
equation (which is relaxational) have all shown \cite{9,10,11,12,13} that
noise selects the wave number which minimizes the underlying free energy, as
expected.  In a series of papers which are particularly relevant to our work,
Kerszberg carried out numerical simulations of directional soldification with
and without Gaussian white noise \cite{14,15}.  This is a system which has no
underlying free energy.  With noise added, the system restabilized into a
cellular state with a unique, reproducible wave number independent of initial
conditions, while without noise the final wave number depended on initial
conditions.

The relaxational, or ``gradient'', case is quite helpful for motivating our
calculations.  Suppose the state of the system is specified by giving a set
of amplitudes $x_k$ of Fourier modes of wave number $k$, chosen so that the
simple steady state of the system is $x_k \equiv 0$, and that the control
parameter has been set so that this state is linearly unstable.  (We will
also take the $x_k$ to be real, so that the pattern is left-right symmetric.)
The amplitudes evolve according to
\begin{equation}
  {dx_k \over dt} \propto -{\partial \Phi(x) \over \partial x_k},
 \label{gradient}
\end{equation}
where $\Phi(x)$ is the free energy.  This evolution always makes $\Phi$
decrease with time.   A patterned steady state with wave number $q$ will have
$x_k$ nonzero only when $k$ is an integer multiple of $q$; let the value of
the free energy of this state be $\Phi_{ss}(q)$.  If we add noise of strength
$\epsilon$ to
(\ref{gradient}), it will produce occasional large fluctuations which take
the system far enough out of the local free energy minimum at the state of
wave number $q$ that it then relaxes to a different local minimum, with wave
number $q'$.  The relative
probability of a transition from $q$ to $q'$ versus a transition from $q'$
back to $q$ is proportional to $\exp[(\Phi_{ss}(q)-\Phi_{ss}(q'))/\epsilon]$.
That is, fluctuations are more likely to take the system from a state of
higher free energy to one of lower free energy than vice versa.  In pattern
forming systems, the steady states near the edges of the band of stable wave
numbers, which are almost unstable, have higher free energies than states in
the interior of the band, so noise drives the system away from those states
and toward the absolute minimum of the free energy.

We argue here that the same thing happens in non-gradient systems, i.e., ones
in which the time evolution is given by
\begin{equation}
  {dx_k \over dt} = F_k(x),
 \label{nonoise}
\end{equation}
but the $F_k$ cannot be put into the form (\ref{gradient}).  To see this, we
first add noise to the dynamics to obtain the Langevin equation
\begin{equation}
  {dx_k \over dt} = F_k(x) + \sqrt{\epsilon} \, \xi_k(t),
 \label{withnoise}
\end{equation}
where the $\xi_k$ are independent Gaussian random variables with mean zero
and unit variance,
\begin{equation}
  \langle \xi_k(t) \, \xi_{k'}(t') \rangle = \delta_{kk'} \, \delta(t-t').
\end{equation}
The $x_k$ are then random variables, and a standard argument \cite{16} leads
to the Fokker-Planck equation, which describes the time evolution of their
probability distribution.  In the long-time limit, this converges to a
steady-state distribution ${\cal P}_{ss}(x)$, which satisfies
\begin{equation}
  -{\partial \over \partial x_k} \left[ F_k(x) {\cal P}_{ss}(x) -
     {\epsilon \over 2} {\partial {\cal P}_{ss} \over \partial x_k} \right]
     = 0.
 \label{Psseqn}
\end{equation}
Here and below, repeated indices are to be summed.  Since the noise terms in
(\ref{withnoise}) are independent of $x$, there is no difference here between
the It\^o and Stratonovich interpretations.

In the weak-noise limit $\epsilon \to 0$, we can solve (\ref{Psseqn}) using a
WKB method \cite{17}.  Since probability distributions must be positive, we
may write
\begin{equation}
  {\cal P}_{ss}(x) \equiv \exp[-S(x)/\epsilon]
 \label{Sdef}
\end{equation}
so that $S$ plays a role similar to that of the free energy in the gradient
case; in fact the solution of (\ref{Psseqn}) is $S = 2\Phi$ in that case.
{}From (\ref{Psseqn}) we then find that to leading order in $\epsilon$, $S$
satisfies
\begin{equation}
  {\partial S \over \partial x_k} \left[ {1 \over 2}
      {\partial S \over \partial x_k} + F_k(x) \right] = 0.
 \label{Seqn}
\end{equation}
{}From this we see that any steady state of the deterministic dynamics
(\ref{nonoise}), i.e., a point where all $F_k$ vanish, is a stationary point
of $S$.  We will see below that a linearly stable steady state of
(\ref{nonoise}) is in fact a local minimum of $S$.

We can also see that $S$ plays the role of $\Phi$ in a different sense.
The deterministic evolution (\ref{nonoise}) never makes $S$ increase, since
\begin{equation}
  {dS \over dt} = {\partial S \over \partial x_k} F_k(x) =
                  - {1 \over 2} {\partial S \over \partial x_k}
                                {\partial S \over \partial x_k} \le 0.
\end{equation}
Thus even a non-gradient system can be considered relaxational, since there
is a function $S(x)$ which is non-increasing.  In gradient systems, however,
that function is just the free energy, which is generally easier to compute.

Eqn. (\ref{Seqn}) can be solved by the method of characteristics, or
equivalently by recognizing it as the Hamilton-Jacobi equation, with energy
0, for the Hamiltonian
\begin{equation}
  {\cal H}(p,x) \equiv -{1 \over 2} p_k p_k - p_k F_k(x).
 \label{Hdef}
\end{equation}
The characteristic curves are given by
\begin{equation}
  \dot x_k = -p_k - F_k(x), \qquad
    \dot p_k = p_{k'} {\partial F_{k'} \over \partial x_k},
 \label{trajectory}
\end{equation}
where overdots represent derivatives with respect to some parameter $\tau$.
Along a characteristic, $S$ is given by
\begin{equation}
  \dot S = p_k \dot x_k = -p_k (p_k + F_k).
 \label{Sdot}
\end{equation}
Since we have ${\cal H} = 0$, this becomes $\dot S = -(p_k p_k)/2$, so $S$
always {\it decreases\/} along a characteristic.

To find the preferred wave number, we need only calculate $S(x)$ at each
stable steady state of (\ref{nonoise}).  The wave number of the state with
the lowest value of $S$ is the preferred wave number:  by (\ref{Sdef}), if
the wave numbers are discrete then all others are exponentially less
probable.  However, since $S$ always decreases along a characteristic, there
are no characteristics which run from one local minimum of $S$ to another.
Thus in order to compare the values of $S$ at its various local minima, we
need characteristics running from one common point to each minimum.
Fortunately, we can find these -- the common point is the unstable steady
state $x_k \equiv 0$, the simple steady state of the system.  That there
should be a characteristic linking this to each stable steady state comes
from a counting argument.  Fixed points of (\ref{trajectory}) occur where
$p_k \equiv 0$ and $F_k(x) \equiv 0$; linearizing about any fixed point gives
\begin{equation}
  \dot {\delta x} = -p - M \delta x, \qquad
    \dot p = M^T p,
 \label{linearized}
\end{equation}
where $M$ is the matrix which gives the linear stability of the deterministic
dynamics (\ref{nonoise}),
\begin{equation}
  M_{kk'} = \left. {\partial F_k \over \partial x_{k'}} \right|_{ss}.
\end{equation}
If there are $N$ amplitudes $x_k$, then the phase space $(x,p)$ is
$2N$-dimensional; from (\ref{linearized}) we see that there are $N$
eigenvectors at each fixed point with eigenvalues which are the same as the
stability eigenvalues of the corresponding deterministic steady state, and
another $N$ eigenvectors with the negatives of those eigenvalues -- and all
$p$ components equal to 0.  From
(\ref{trajectory}), we see that the $p = 0$ subspace of phase space is
invariant, and (\ref{Sdot}) shows that following a trajectory in this
subspace would leave $S$ unchanged.  Thus we want characteristics {\it not\/}
to be in this subspace.  However, the only eigenvectors coming out from a
linearly stable steady state (i.e., having positive eigenvalues) are those
which lie in the $p=0$ subspace.  Thus we cannot have a characteristic
leaving a stable steady state.  All nearby characteristics
must then be directed toward that state.  Since $S$ always decreases
along characteristics, a linearly stable steady state of the deterministic
dynamics must then be a local minimum of $S$.  However, the steady state at
$x=0$ is not stable; it has one linearly unstable direction for each
fundamental wave number $q$ against which it is unstable.  Correspondingly,
for each such $q$ there is a steady state which bifurcated from
the $x=0$ state as the control parameter was increased to its present value.
It is then natural to expect that there will be one characteristic running
from $(x,p)=(0,0)$ to each linearly stable steady state (and $p=0$).

To do the calculation numerically, it is useful to have the expression for
the Lagrangian corresponding to the Hamiltonian (\ref{Hdef}):
\begin{equation}
  {\cal L}(x,\dot x) = -{1 \over 2} [\dot x_k + F_k(x)] [\dot x_k + F_k(x)]
\end{equation}
{}From (10), we see that on the characteristics this is equal to
$-(p_k p_k)/2$, which in turn is equal to $\dot S$.  Thus we can set up the
calculation of $S$ at the steady state as a minimization problem, using
Hamilton's principle:  we wish to find the trajectory which approaches
$(x,p)=(0,0)$ as $\tau \to -\infty$, approaches the steady state $x$ (and
$p=0$) as $\tau \to \infty$, and minimizes $\int {\cal L} \, d\tau$.  The
resulting minimum value is the value of $S$ for that particular steady state.
Minimizing this result in turn over the possible fundamental wave numbers $q$
gives the preferred wave number.  This procedure is made much easier by the
fact that the structure of the deterministic equations (\ref{nonoise}) is
usually such that the multiples of any fundamental wave number $q$ form an
invariant subspace.  This property is inherited by the equations
(\ref{trajectory}) for the characteristics, and so only a relatively small
number of modes must be kept for each $q$.

When the control parameter is only slightly beyond the onset of instability
of the simple steady state, it is possible to calculate the preferred wave
number analytically.  The simplest nontrivial example of this procedure is
for a system in which a symmetry keeps the $k=0$ amplitude fixed.  Expanding
to third order in the interface displacement generally leads to equations of
the form (\ref{nonoise}) with
\begin{eqnarray}
    &&F_q = \sigma_q x_q - \lambda x_q^3 - \mu x_q x_{2q}, \nonumber \\
    &&F_{2q} = -|\sigma_{2q}| x_{2q} + \alpha x_q^2, \nonumber \\
    &&F_{3q} = -|\sigma_{3q}| x_{3q} + \beta x_q^3 + \gamma x_q x_{2q}.
  \label{exampleF}
\end{eqnarray}
All coefficients depend on the fundamental wave number $q$, and near the
onset of instability of the state $x \equiv 0$ the linear growth rate
$\sigma_q$ is small and positive.  The $p_{3q}$ equation coming from
(\ref{trajectory}) immediately gives
$p_{3q} \propto \exp(-|\sigma_{3q}|\tau)$, and since $p_{3q}$ must go to zero
for $\tau \to -\infty$ this forces $p_{3q} \equiv 0$.  This makes $x_{3q}$
irrelevant, since it does not appear in $F_q$ or $F_{2q}$, and in $\dot S$ it
is multiplied by $p_{3q}$.  The patterned steady state is given by
\begin{eqnarray}
    x_q^{ss} &&= [\sigma_q |\sigma_{2q}| /
                  (\lambda |\sigma_{2q}| + \alpha \mu)]^{1/2}
     + O(\sigma_q^{3/2}), \nonumber \\
    x_{2q}^{ss} &&= \sigma_q \alpha / (\lambda |\sigma_{2q}| + \alpha \mu)
     + O(\sigma_q^2),
\end{eqnarray}
with the corrections coming from higher order terms which were neglected in
writing Eqn. (\ref{exampleF}).  With the rescalings
\begin{equation}
  x_q = \sigma_q^{1/2} x, \quad p_q = \sigma_q^{3/2} p, \quad
    x_{2q} = \sigma_q y, \quad p_{2q} = \sigma_q^2 r,
\end{equation}
we find that $\dot x$ and $\dot p$ are explicitly of order $\sigma_q$, while
to leading order we always have
\begin{equation}
  y = \alpha x^2 / |\sigma_{2q}|, \qquad r = -\mu p / |\sigma_{2q}|.
\end{equation}
Substituting all this into the expression (\ref{Hdef}) for the Hamiltonian
and setting it equal to zero then gives
\begin{equation}
  p = -2 x + 2 (\lambda |\sigma_{2q}| + \alpha \mu) x^3 / |\sigma_{2q}|.
\end{equation}
It is then simple to integrate (\ref{Sdot}) to leading order to get the value
of $S$ at the steady state:
\begin{equation}
  S(x^{ss}) = - \sigma_q^2 |\sigma_{2q}| / 2(\lambda |\sigma_{2q}|
                + \alpha \mu).
  \label{exampleS}
\end{equation}
To find the preferred wave number $q$ to leading order, we must
minimize (\ref{exampleS}) over all $q$ in the unstable band.  Higher-order
corrections can also be calculated, although to do this properly we need to
retain higher-order corrections in (\ref{exampleF}), which in turn requires
us to keep more than three Fourier modes.  Note that in the special case
$\mu = -2\alpha$, (\ref{exampleF}) is a gradient system (to the relevant
order in $\sigma_q$).  However, this leads to no simplification of the
calculation of $S$, although the result then agrees with the free energy.

We have carried out the analogous calculations for the Greenside-Cross
equation \cite{18} for B\'enard convection, which is a non-gradient equation
for a spatial order parameter $\psi$,
\begin{equation}
  \partial \psi / \partial t = [\gamma - (\nabla^2 + 1)^2] \psi
                               + 3 |\nabla\psi|^2 \nabla^2\psi.
\end{equation}
The steady state $\psi \equiv 0$ is linearly unstable for $\gamma>0$.
Keeping modes up to the seventh harmonic, we find that the preferred wave
number for small $\gamma$ is given by
\begin{equation}
  q = 1 - {\gamma \over 4} - {101 \gamma^2 \over 1024} -
            {981 \gamma^3 \over 16384} - O(\gamma^4).
 \label{qseries}
\end{equation}
In {\it one\/} spatial dimension, the Greenside-Cross equation is a gradient
system, and this result agrees with the expansion obtained by minimizing the
free energy.  Cross and Meiron \cite{13}, in numerical simulations of
deterministic, two-dimensional evolution with random initial conditions, find
that the system reaches a wave number $q \approx 0.78$ for long times for
$\gamma = 1/2$.  Setting $\gamma=1/2$ in (\ref{qseries}) gives $q=0.843$,
although the series has clearly not converged (the first three terms give
0.850).

In this Letter we have argued that pattern forming systems have a naturally
preferred wave number, namely that which the system would approach if it
were subjected to weak, additive, Markovian, Gaussian white noise.  This is
not to say that the noise actually experienced by such a system has any of
these five properties; but Gaussian white noise is appropriate because it
does not bias the system's preference for a wave number, as colored noise,
for example, would do.  The argument does suggest, however, that the role of
noise is not limited to providing the initial fluctuation which takes the
system out of its simple steady state and starts it evolving toward one of
its possible restabilized states, as is usually taken (implicitly) to be the
case.  Rather it has a continuing role in readjusting the wave number of the
patterned state, generally through occasional large fluctuations which either
create or destroy a cell.

An obvious question which needs to be addressed in the future is that of the
rate at which the steady-state probability distribution ${\cal P}_{ss}$ is
established.  It should be possible to use the same classical-mechanics
techniques to study this question as we have used above, since substituting
the time-dependent generalization of ansatz (\ref{Sdef}) into the
Fokker-Planck equation leads directly to the time-dependent Hamilton-Jacobi
equation for the Hamiltonian (\ref{Hdef}).  Further work on this point is
under way.

Other important issues arise when we model the noise to which a system is
actually subjected, which may not be additive, white, Gaussian, Markovian, or
weak.  For instance, the relevant fluctuations might be in the value of the
control parameter.  If any of the first four properties are lacking, then the
appropriate Fokker-Planck equation will not have the simple form
(\ref{Psseqn}).  Even if the relevant noise source is thermal, evolution
equations of the form (\ref{nonoise}) often arise only after considerable
manipulation of some more complex model which is written down from first
principles.  The $\epsilon$ in (\ref{withnoise}) may then be replaced by
something which depends on $q$ and even $x$.  In all such cases the
subsequent calculations need to be modified appropriately.

When the noise strength is finite, several important effects arise.  One is
that we may need the next higher order correction to the leading-order $S(x)$
which we have calculated.  A second, related point is that the relevant
probabilities to compare are not just the heights of the peaks in
${\cal P}_{ss}$, but the areas under the peaks -- including not just the
exact steady states, but also perturbations of those states.  These effects
have been seen by Kerszberg \cite{19}, who found that even in a gradient
system the observed wave number was not equal to the wave number which
minimized the free energy when noise was included in his calculations.
Finally, in an infinite system the possible band of wave numbers is
continuous, so that it is not true that one wave number has a probability
which is exponentially larger than all others.  Rather the probability
$\exp[-S(q)/\epsilon]$ is appreciable for all wave numbers within a range of
order $\epsilon^{1/2}$ around the ``preferred'' wave number.  Thus if the
noise variance is larger than something of the order of the inverse square of
the size of the system, then there is still a range of possible wave numbers
that one might observe even in the long time limit, albeit a very narrow
range for finite but weak noise.

Finally, it would of course be far more satisfying to have a direct physical
interpretation of the criterion for the preferred wave number, rather than
just a prescription for calculating it.

This work was supported by the National Aeronautics and Space Administration,
through grant number NAG3-1603 and also through the JOVE program at North
Dakota State University.


\begin{references}

\bibitem[*]{email} Electronic address:  kurtze@plains.NoDak.edu

\bibitem{1} A.C. Newell, C.G. Lange, P.J Aucoin, and J.F Mack,
J. Fluid Mech. {\bf 40}, 513 (1970).

\bibitem{2} H.R Schober, E. Allroth, K. Schroeder, and
H. M\"uller-Krumbhaar, Phys. Rev. A {\bf 33}, 567 (1986).

\bibitem{3} C.W. Meyer, G. Ahlers, and D.S. Cannell, Phys. Rev. Lett.
{\bf 59}, 1577 (1987).

\bibitem{4} P.C. Hohenberg and J.B. Swift, Phys. Rev. A {\bf 46}, 4773
(1992).

\bibitem{5} R. Pieters and J.S. Langer, Phys. Rev. Lett. {\bf 56}, 1948
(1986).

\bibitem{6} J.S. Langer, Phys. Rev. A {\bf 36}, 3350 (1987).

\bibitem{7} X.W. Qian, H. Chou, M. Muschol, and H.Z. Cummins, Phys. Rev. B
{\bf 39}, 2529 (1989).

\bibitem{8} J.A. Warren and J.S. Langer, Phys. Rev. E {\bf 47}, 2702
(1993).

\bibitem{9} J. Vi\~nals, E. Hern\'andez-Garc\'{\i}a, M. San Miguel, and
R. Toral, Phys. Rev. A {\bf 44}, 1123 (1991).

\bibitem{10} H.W. Xi, J. Vi\~nals, and J.D. Gunton, Physica A {\bf 177},
356 (1991).

\bibitem{11} K.R. Elder, J. Vi\~nals, and M. Grant, Phys. Rev. Lett.
{\bf 68}, 3024 (1992).

\bibitem{12} K.R. Elder, J. Vi\~nals, and M. Grant, Phys. Rev. A {\bf 46},
7618 (1992).

\bibitem{13} M.C. Cross and D.I. Meiron, Phys. Rev. Lett. {\bf 75}, 2152
(1995).

\bibitem{14} M. Kerszberg, Phys. Rev. B {\bf 27}, 3909 (1983).

\bibitem{15} M. Kerszberg, Phys. Rev. B {\bf 28}, 247 (1983).

\bibitem{16} See, e.g., R.L. Stratonovich, {\it Introduction to the Theory
of Random Noise}, (Gordon and Breach, 1963), vol. 1.

\bibitem{17} See, e.g., R.L. Stratonovich in {\it Noise in Nonlinear
Dynamical Systems}, edited by F. Moss and P.V.E. McClintock (Cambridge,
1989), vol. 1, p. 16.

\bibitem{18} H.S. Greenside and M.C. Cross, Phys. Rev. A {\bf 31}, 2492
(1985).

\bibitem{19} M. Kerszberg, Phys. Rev. A {\bf 28}, 1198 (1983).

\end{references}
\end{document}